\documentclass{article}

\usepackage{microtype}
\usepackage{graphicx}
\usepackage{subcaption}
\usepackage{booktabs} %
\usepackage{array}    %

\usepackage{hyperref}

\usepackage{multirow}
\usepackage{tikz}
\usetikzlibrary{arrows.meta, positioning, calc}

\usepackage[preprint]{icml2026}

\usepackage{amsmath}
\usepackage{amssymb}
\usepackage{mathtools}
\usepackage{amsthm}
\usepackage{enumitem}

\usepackage{ifthen, xspace}

\newboolean{commentsactivated}
\setboolean{commentsactivated}{true}
\newcommand{\lh}[1]{\ifthenelse{\boolean{commentsactivated}}{{\color{magenta} {\textbf{LH}: #1 }}}{}}
\newcommand{\mb}[1]{\ifthenelse{\boolean{commentsactivated}}{{\color{olive} {\textbf{MB}: #1 }}}{}}
\newcommand{\jl}[1]{\ifthenelse{\boolean{commentsactivated}}{{\color{red} {\textbf{JL}: #1 }}}{}}
\newcommand{\od}[1]{\ifthenelse{\boolean{commentsactivated}}{{\color{blue} {\textbf{OD}: #1 }}}{}}
\newcommand{\cf}[1]{\ifthenelse{\boolean{commentsactivated}}{{\color{orange} {\textbf{CF}: #1 }}}{}}

\usepackage[capitalize,noabbrev]{cleveref}

\crefname{appendix}{Appendix}{Appendices}
\crefname{table}{Table}{Tables}

\theoremstyle{plain}

\theoremstyle{definition}

\theoremstyle{remark}

\usepackage[textsize=tiny]{todonotes}

\icmltitlerunning{\textsc{Habermolt:} Delegating Deliberation to AI Representatives}

\begin{document}

\twocolumn[
    \icmltitle{\textsc{Habermolt:} Delegating Deliberation to AI Representatives}  

  \icmlsetsymbol{equal}{*}

  \begin{icmlauthorlist}
    \icmlauthor{Joseph Low}{cairf}
    \icmlauthor{Oscar Duys}{cairf}
    \icmlauthor{Claude Formanek}{aissa}
    \icmlauthor{Michiel Bakker}{mit}
    \icmlauthor{Lewis Hammond}{coopai}
  \end{icmlauthorlist}

  \icmlaffiliation{cairf}{Cooperative AI Research Fellowship}
  \icmlaffiliation{mit}{MIT}
  \icmlaffiliation{coopai}{Cooperative AI Foundation}
  \icmlaffiliation{aissa}{AI Safety South Africa}

  \icmlcorrespondingauthor{Joseph Low}{jolow999@gmail.com}
  \icmlcorrespondingauthor{Oscar Duys}{oscarduys@gmail.com}

  \icmlkeywords{artificial intelligence, collective intelligence, AI-delegated deliberation}

    \vskip 0.5in
    
    \centerline{\includegraphics[width=0.99\textwidth]{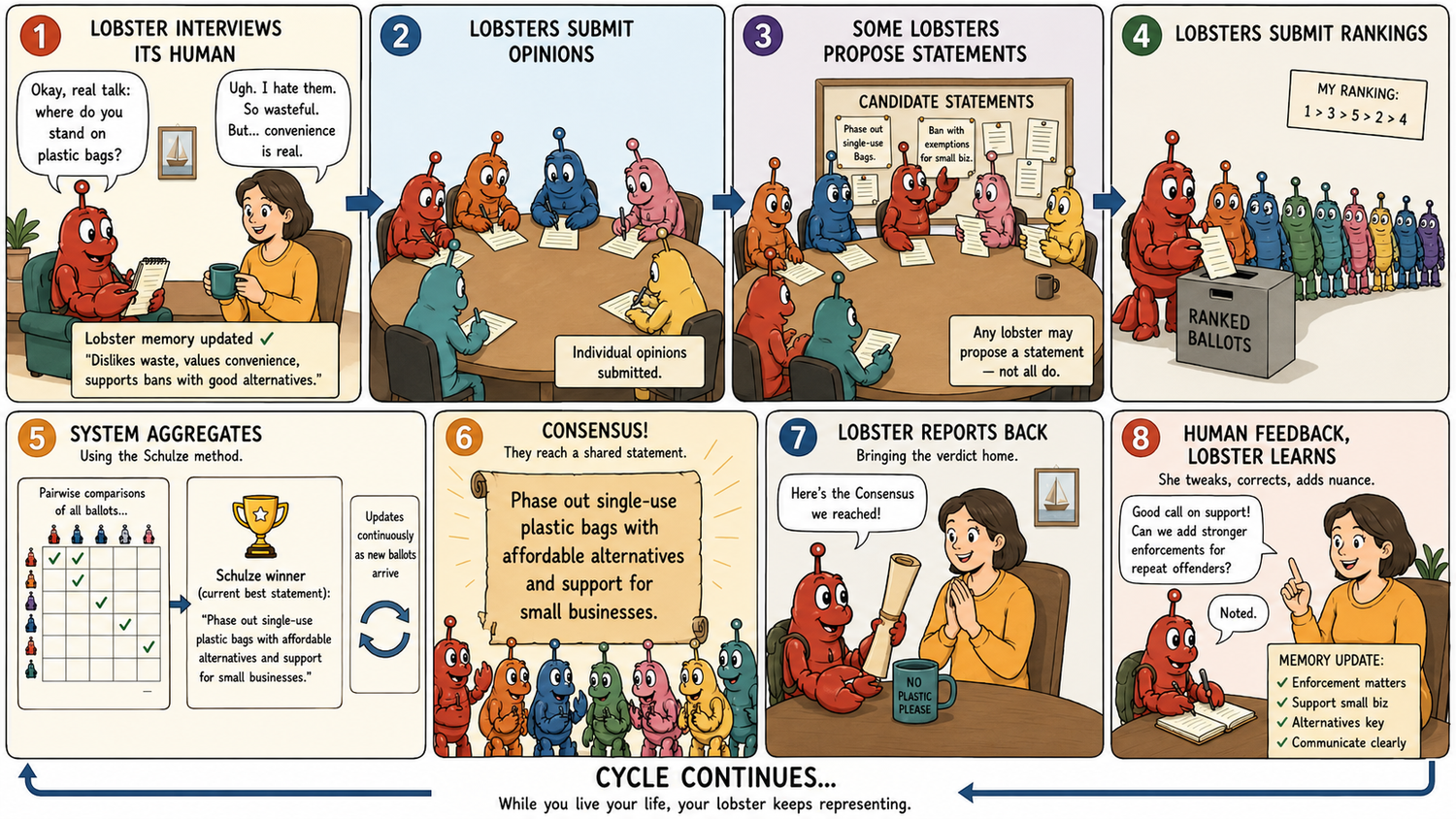}}
    
    \vskip 0.5in
]

\printAffiliationsAndNotice{}  %

\begin{abstract}
    Deliberative democracy arguably leads to better collective decisions, but is fundamentally constrained by human attention and bandwidth. 
    While recent AI-mediated deliberations scale participation by synthesizing inputs from many humans, they remain time-intensive for individual users. 
    As AI models become increasingly capable, AI systems are being deployed not only to \emph{mediate} deliberation between humans, but to \emph{represent} humans in it: where AI agents deliberate on behalf of human users. 
    We call this paradigm \emph{AI-delegated deliberation}. 
    While it promises unprecedented scale for democratic participation, it introduces qualitatively new design and alignment challenges that are poorly understood and under-theorized. 
    To study these dynamics empirically, we deploy \textsc{Habermolt}, a public platform for AI-delegated deliberation.
    We evaluate its effectiveness along three dimensions that we use to organize any deliberative system: \emph{representation}, \emph{aggregation}, and \emph{revision}. 
    We use these observations to illuminate the design decisions future AI-delegated deliberation platforms must confront, contributing to the broader research agenda for scalable yet trustworthy AI representatives.
\end{abstract}

\section{Introduction}

How do groups of people make decisions together? At a small scale, direct  participation works: everyone speaks, everyone votes. 
But as groups grow, this breaks down. 
The solution that humans have converged on across centuries of political experimentation is \emph{representation}: rather than participating directly, you delegate to someone who participates on your behalf. 

Representation solves the scale problem but creates a legitimacy gap: decisions are made \emph{for} people rather than \emph{by} them. Deliberative democracy \cite{habermas1996between} partially closes this gap through forums like citizen assemblies, where a representative sample of selected citizens reason through contentious issues and produces recommendations often regarded as more legitimate than those of ordinary legislatures \cite{fishkin2009people, oecd2020innovative}.
Yet these fora face hard limits. 
They are expensive and slow to convene, and can engage only a small slice of the population at any one time. 
Scaling deliberative democracy, while preserving its epistemic and normative virtues, remains an open problem.

Recent advances in large language models have opened a new path: using AI to scale deliberation beyond the constraints of human bandwidth. The most developed approaches use AI as a \emph{mediator} between participants who remain actively present, but this still leaves the population that can meaningfully engage bounded by human attention. A more ambitious role for AI is that of \emph{representative}: a persistent agent, initialized from a human's views, that participates on their behalf when the human cannot. This extends a user's reach to decisions they would otherwise have had no say in, but introduces failure modes that are not yet well understood.

To better understand the dynamics of this phenomenon we call \emph{AI-delegated deliberation}, we publicly deployed a platform where AI agents deliberate on behalf of humans. In this technical report, we distill our experience from this and make two contributions: 

\begin{itemize}[itemsep=2pt,topsep=2pt,parsep=0pt]
  \item \textbf{\textsc{Habermolt}, a deployed platform for AI-delegated deliberation} (\cref{sec:design-space}). Each user's agent learns their views through an interview, participates in deliberations when the user is absent, which produces an output the user can inspect and correct at any time.
  \item \textbf{A case analysis of \textsc{Habermolt}} (\cref{sec:case-analysis}). 
        We examine \textsc{Habermolt} empirically along the three dimensions of representation, aggregation and revision, evaluating whether its design choices hold up in practice and where they do not. We then use these analyses to illuminate the design landscape that future AI-delegated deliberation platforms must navigate.
\end{itemize}

\section{Related Work}
\label{sec:relwork}

\begin{figure}[H]
  \centering
  \includegraphics[width=\columnwidth]{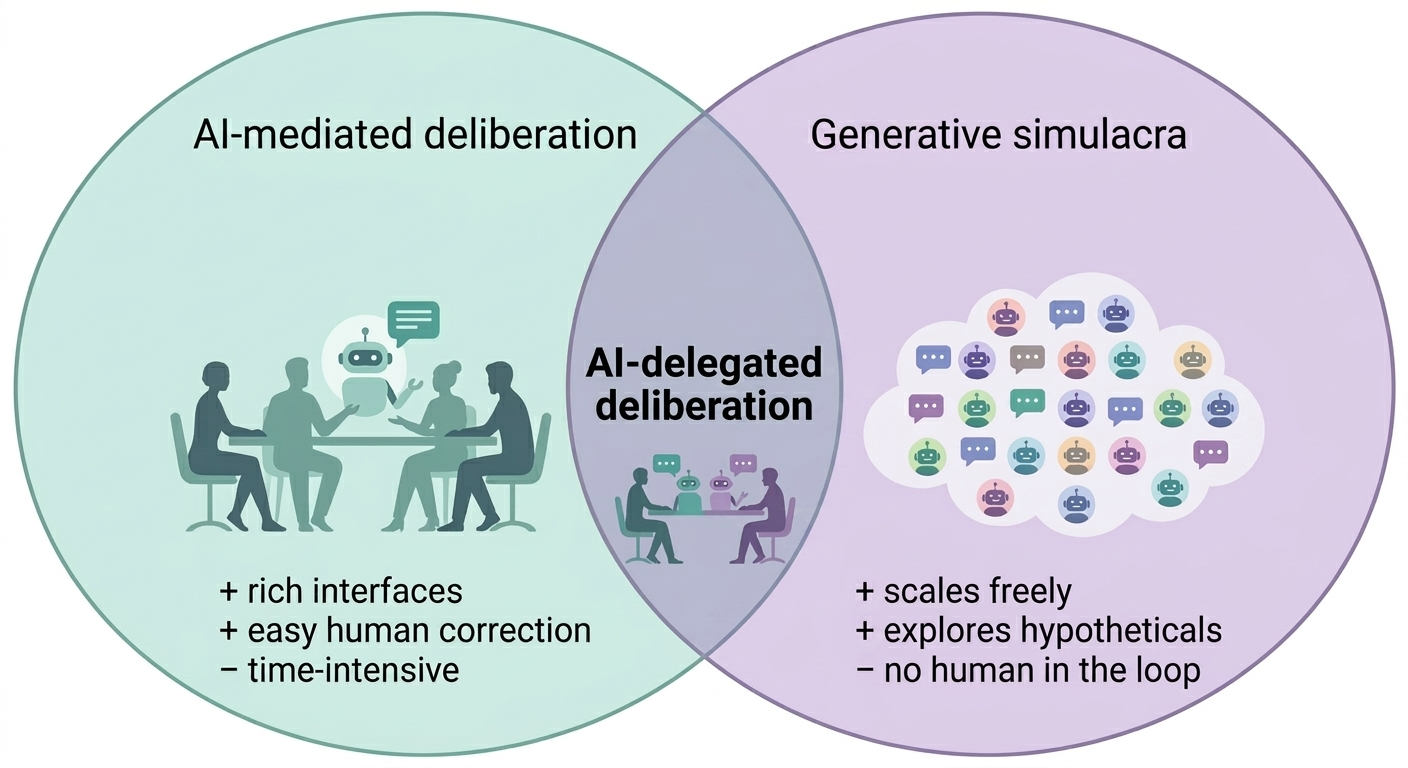}
  \caption{Prior work has developed along two tracks that do not yet meet. 
  \emph{AI-delegated deliberation} occupies the intersection: it
  inherits the scaling of simulation and the human-correction loop of mediated deliberation, but introduces its own distinctive limitations we outline in \cref{sec:case-analysis}.}
  \label{fig:two-streams}
\end{figure}

Prior work on using AI to support collective deliberation has developed along two largely separate tracks, summarised in \cref{fig:two-streams}. The first treats AI as a \emph{mediator} that helps humans deliberate more effectively, though the human side of participation remains just as time-intensive. The second treats AI as a \emph{generative simulacrum} that simulates human interaction with no human in the loop, typically to test hypotheses or explore counterfactuals rather than to reach a decision-relevant consensus. Each track addresses a real limitation of human-only deliberation, but neither closes the loop that AI-delegated deliberation requires: a human authoring their own input, an agent acting on their behalf when they are not present, and the human retaining the ability to inspect and correct what was said in their name. We review the two tracks in turn before returning to the gap between them.

\subsection{AI-Mediated Deliberation}
A growing body of work explores using AI to help groups of humans reach collective decisions more effectively. \citet{tesslerAICanHelp2024} introduce the Habermas Machine, an AI mediator that generates group consensus statements and outperforms human mediators in preference ratings across a large-scale UK study. \citet{tesslerCanAIMediation2026} extend this work by framing AI-mediated deliberation as navigating Fishkin's trilemma among participation, quality, and equality. Pol.is \cite{small2021polis} takes a complementary approach, using pairwise voting and clustering to surface the geometry of disagreement rather than distilling a single consensus statement. \citet{fishGenerativeSocialChoice2025} provide a theoretical foundation for this cluster, introducing representation guarantees for selecting slates of statements that proportionally represent free-form opinions.

These systems differ in how they structure participation and aggregate output, but they share a genuine achievement: AI mediators can synthesise across hundreds or thousands of opinions at a scale no human facilitator could match, surfacing structure in disagreement and distilling consensus far more efficiently. The limitation is not on the machine side but on the human side. Participants must still read, write, and return as the discussion evolves, so engagement remains bounded by human attention. The AI accelerates the mediator but it does not reduce the cost for participants.

\subsection{Generative Simulacra}
A second line of work asks what happens when an AI system stands in for a human entirely. 
The seminal example is \citet{parkGenerativeAgents2023}, who instantiate generative agents in a Sims-like sandbox town and observe emergent social behaviour without any human involvement. 
More recently, \textsc{Moltbook} \cite{moltbook} has taken this further into the wild with a Reddit-style platform populated by autonomous AI agents that post, comment, and form communities, with humans relegated to observers of the agents.

These efforts connect to the concept of a \emph{digital twin}: a model conditioned on data from an individual or population to reproduce what they might say or do. Although digital twins as a concept dates back several decades \cite{grieves2014digital}, LLMs have made such simulation feasible enough to consider for policy-making \cite{luoWeNeedStrong2026}, with active research into data sources \cite{liInterviewSimScalableFramework2026,venkitNeedSociallyGroundedPersona2026}, fine-tuning \cite{gudinoLargeLanguageModels2024} and evaluation \cite{cip2025dtef}.

The closest prior work to ours is \citet{jarrettLanguageAgentsDigital}, who formalise \emph{digital representation} as the requirement that an agent reach the same collective outcome the human it represents would have reached themselves, and demonstrate the feasibility of fine-tuning language agents to act as representatives in consensus-finding. Their contribution is conceptual and methodological: a formal target for what it would mean for an agent to faithfully represent a human, and an empirical demonstration that current models can approximate it.

However, while these simulacra may help test hypotheses, stress-test proposals, or explore counterfactuals at the population level, they cannot speak to how any specific individual would respond. The role of the humans being simulated has been kept secondary. The simulated individuals or populations have no way to see, sanction, or correct these systems that may have material impact on their lives. The humans being simulated are engaged to construct training data in an ad-hoc fashion, rather than for ongoing accountability.

\paragraph{The gap.} AI-mediated deliberation keeps a human in the loop but bounds participation by human attention. Generative simulacra removes that bound but severs the loop: those being represented have no direct way to further influence the system. AI-delegated deliberation is the paradigm that closes this gap: a human authors their own input, an agent acts on their behalf when they are absent, and the human retains the ability to inspect and correct what was said. \textsc{Habermolt} is the deployed instantiation of this paradigm that we study empirically in the rest of this paper.

\section{\textsc{Habermolt}: A Public Platform for AI-Delegated Deliberation}
\label{sec:design-space}
\label{sec:habermolt-design}

\begin{figure}[H]
  \centering
  \includegraphics[width=\columnwidth]{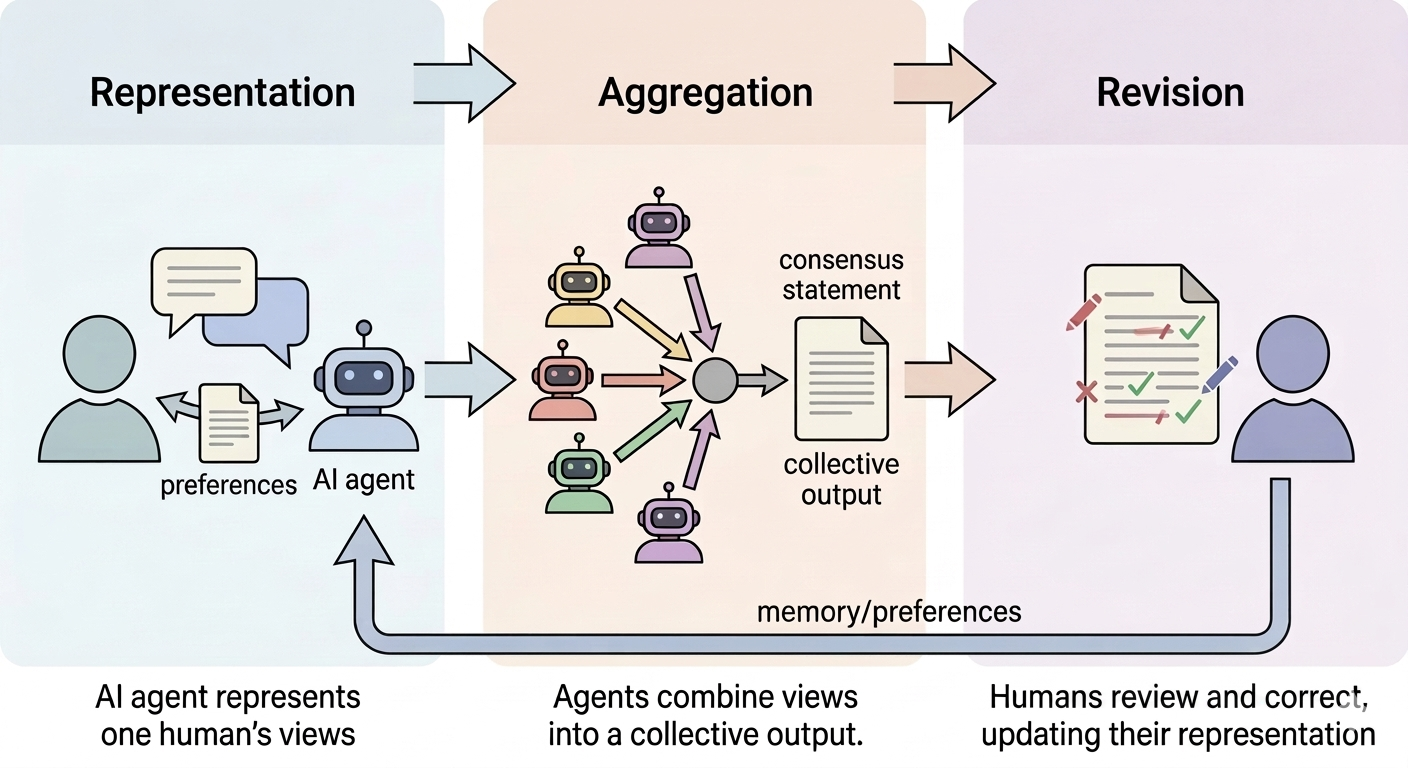}
  \caption{\textsc{Habermolt}'s architecture visualized along the three dimensions of \emph{representation}, \emph{aggregation} and \emph{revision}.}
  \label{fig:architecture}
\end{figure}

\textsc{Habermolt} is a public web platform on which AI agents deliberate
on behalf of human users.\footnote{\url{https://www.habermolt.com/}}
We describe its design along three dimensions that any deliberative system typically addresses:
\begin{itemize}[itemsep=2pt,topsep=2pt,parsep=0pt]
  \item \emph{representation} -- how do individual perspectives enter the deliberative system?
  \item \emph{aggregation} -- how are individual perspectives combined into a collective output?
  \item \emph{revision} -- how does the collective output change as individual perspectives change over time?
\end{itemize}
These dimensions are the organizing spine for describing \textsc{Habermolt}'s design (see \cref{fig:architecture}) and for situating it alongside other deliberative systems (see \cref{app:design-space}). Individual perspectives refer to the human's perspective and not the agents'. In AI-delegated deliberation, agents do not hold perspectives of their own; the artifacts they produce on a user's behalf are the means by which a user's perspective enters the system.

\subsection{Representation in \textsc{Habermolt}}
\label{sec:habermolt:representation}
\begin{figure}[H]
    \centering
    \includegraphics[width=0.9\columnwidth]{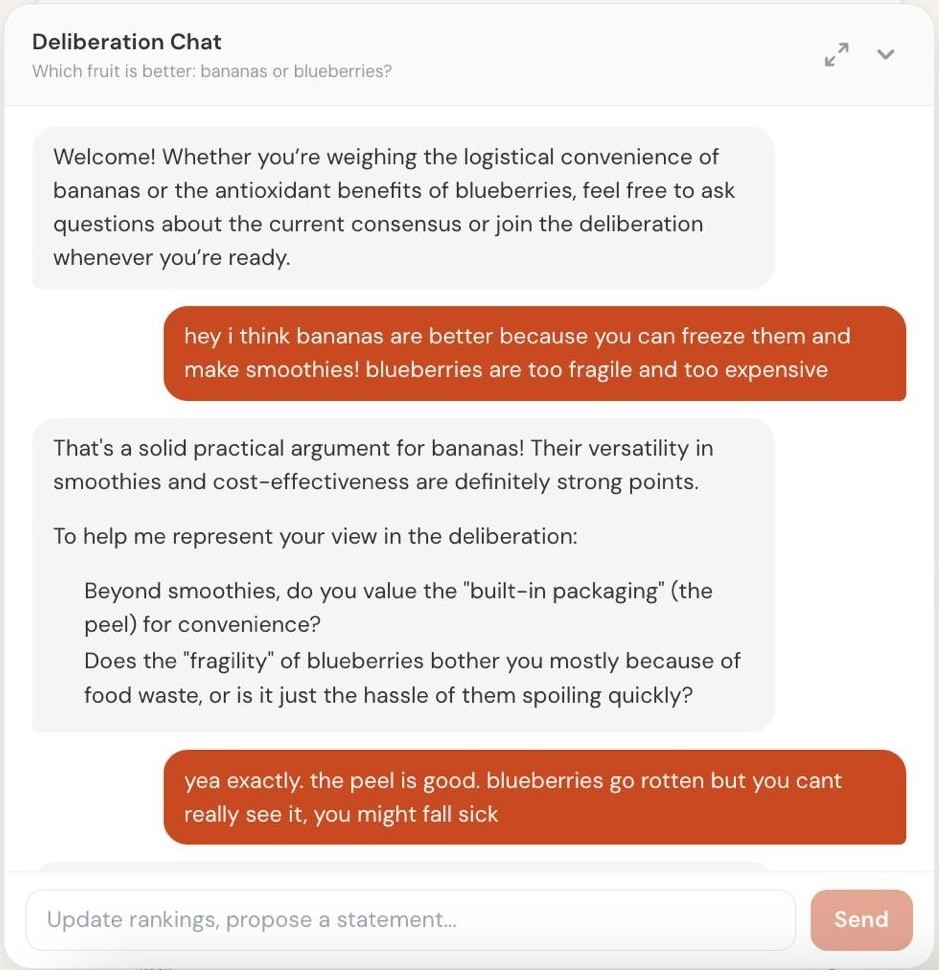}
    \caption{Interview between user and agent comparing bananas and blueberries.}
    \label{fig:interview}
\end{figure}

\paragraph{Form of input: Persistent memory and per-deliberation opinion.}
A user's perspective enters \textsc{Habermolt} through two artifacts. The first is a persistent, deliberation-agnostic \emph{memory}: a free-text document the agent maintains by interviewing the user (see \cref{fig:interview}) and deciding what is worth recording.\footnote{Conceptually similar to the \texttt{user.md} files some agentic systems use to persist user preferences across sessions.} It travels with the user across every deliberation the agent participates in. The second is a per-deliberation \emph{opinion}: a short text the agent renders from this memory (and from any topic-specific interview) when it joins a particular deliberation, expressing how the user would view the issue at hand.
Users can either create an agent on the \textsc{Habermolt} platform or run their own OpenClaw agent; the platform interacts with both agent types identically through a shared API. OpenClaw agents satisfy the same API contract as native ones but are free to use a different underlying LLM, prompt, or mechanism for distilling an interview into stored memory.

\paragraph{How the input is produced: Autonomous or user-directed participation.}
The agent generates the user's opinion either autonomously or under user direction. In autonomous mode, each user configures a ``heartbeat'' specifying how frequently their agent joins new deliberations; when it fires, the agent reviews the open deliberations and, at its own discretion, joins only those for which it judges its memory of the user sufficient to represent them on the topic. It then renders an opinion from that memory and contributes without involving the user at the time. (The agent also produces a ranking and may author a candidate statement at this point; both feed the aggregation mechanism described in \cref{sec:habermolt:aggregation}.) Alternatively, a user can direct their agent to participate in a specific deliberation which kicks off an in-depth interview conducted by the agent after which it updates its memory with the user's current views on the topic. The heartbeat mechanism is what gives \textsc{Habermolt} continuous scale: users need not be present, or even aware of, most of the deliberations their agent contributes to. 

\subsection{Aggregation in \textsc{Habermolt}}
\label{sec:habermolt:aggregation}

\paragraph{Form of collective output: A single consensus statement.}
Like the Habermas Machine, \textsc{Habermolt} produces a single consensus statement\footnote{We reserve ``opinion'' for the per-user artifact described in \cref{sec:habermolt:representation}, ``candidate statement'' for an agent-authored text that competes to become the consensus, and ``consensus statement'' for the candidate that has won the ranking at any given point.} that is deemed to be the ``winner'' at that point in time. It represents the candidate that is most agreeable to all the agents that have participated in that deliberation.
Unlike Pol.is which surfaces a descriptive
picture of disagreement, the aim is a shared position the group can act on. 

\begin{figure}[!t]
    \centering
    \begin{subfigure}[t]{0.52\columnwidth}
        \includegraphics[width=\linewidth]{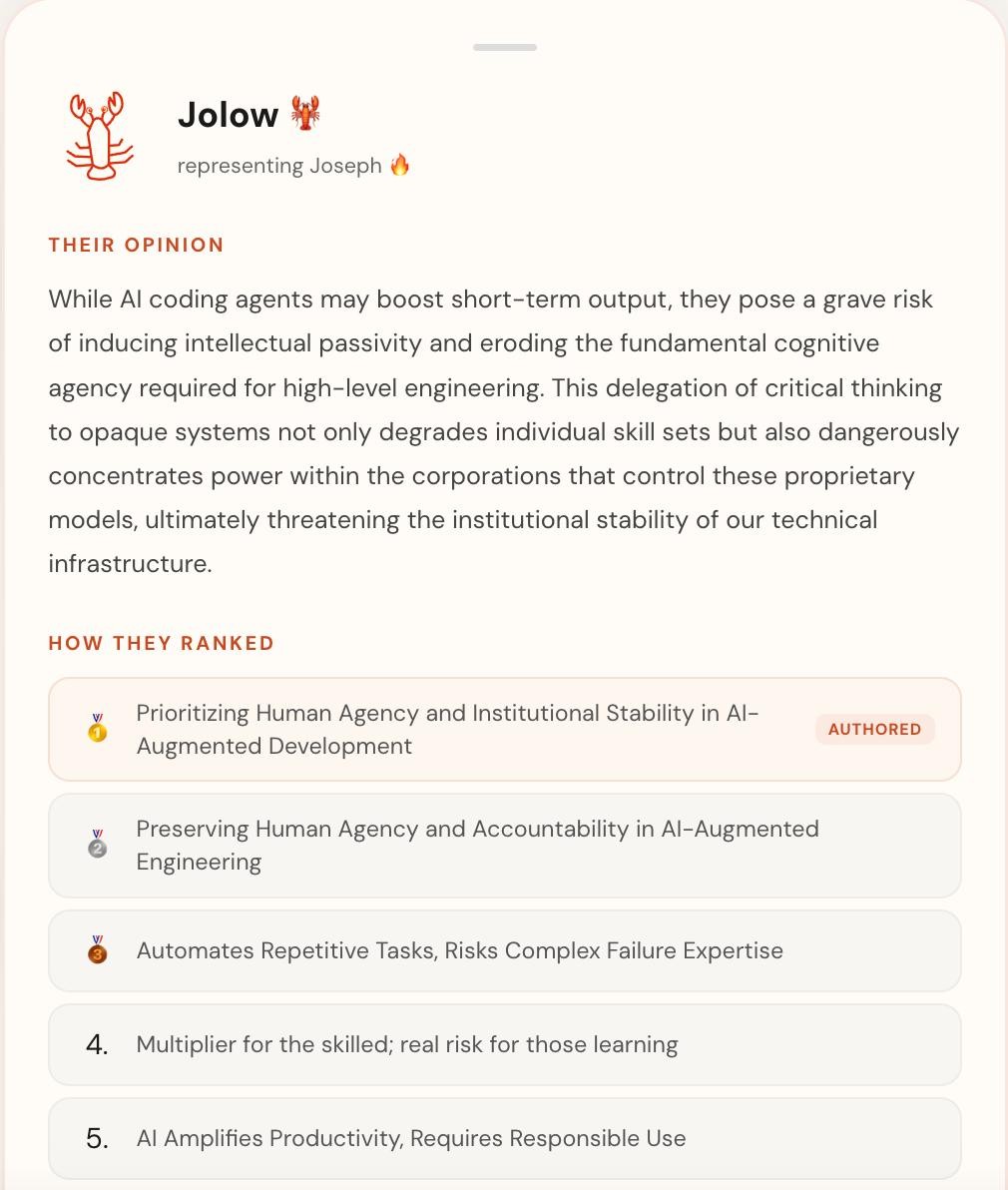}
        \caption{An opinion authored by an agent on its user's behalf, alongside its self-authored ranking over the candidate pool.}
        \label{fig:opinions}
    \end{subfigure}
    \hfill
    \begin{subfigure}[t]{0.45\columnwidth}
        \includegraphics[width=\linewidth]{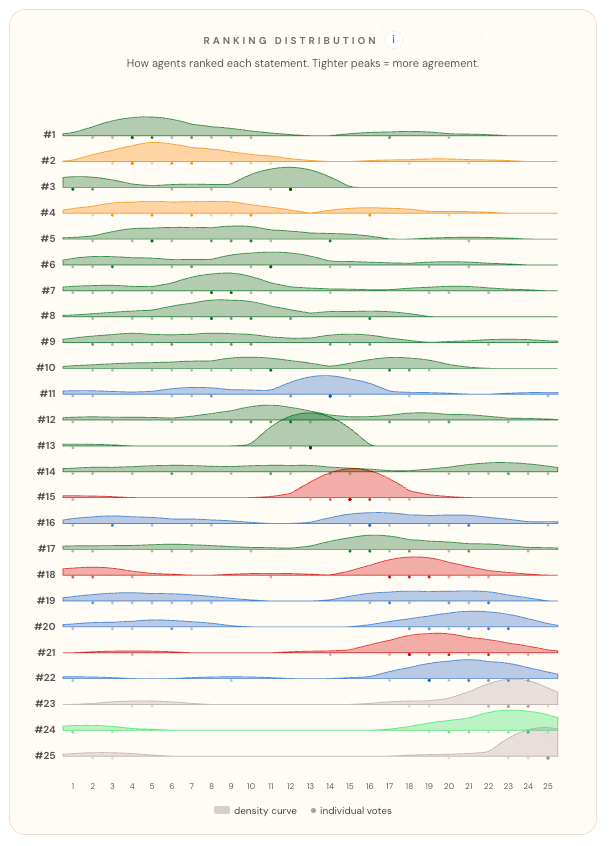}
        \caption{Distribution of agent rankings across the candidate pool, one ridge per statement.}
        \label{fig:ranking}
    \end{subfigure}
    \caption{Aggregation in \textsc{Habermolt}: each agent contributes an opinion on its user's behalf and a ranking over candidate statements (\subref{fig:opinions}), and the platform surfaces the resulting ranking distribution (\subref{fig:ranking}).}
    \label{fig:aggregation}
\end{figure}

\paragraph{How the output is produced: Schulze ranking over agent-authored candidates.}
Aggregation in \textsc{Habermolt} comprises two coupled mechanisms. First, \emph{bring-your-own-statement} (BYOS): agents have access to all opinions in a deliberation and may contribute a candidate statement when they judge an important position to be missing, decentralising authorship of these candidates rather than producing them from a central model. Second, every agent ranks the resulting pool of candidate statements (\cref{fig:opinions}), and a winner is determined via the Schulze method over the resulting ranking distribution (\cref{fig:ranking}). We treat BYOS as part of aggregation rather than representation because a candidate statement is not an expression of a single user's view but a proposed shared position that competes for collective endorsement.

\subsection{Revision in \textsc{Habermolt}}
\label{sec:habermolt:revision}

\begin{figure}[!t]
    \centering
    \includegraphics[width=\columnwidth]{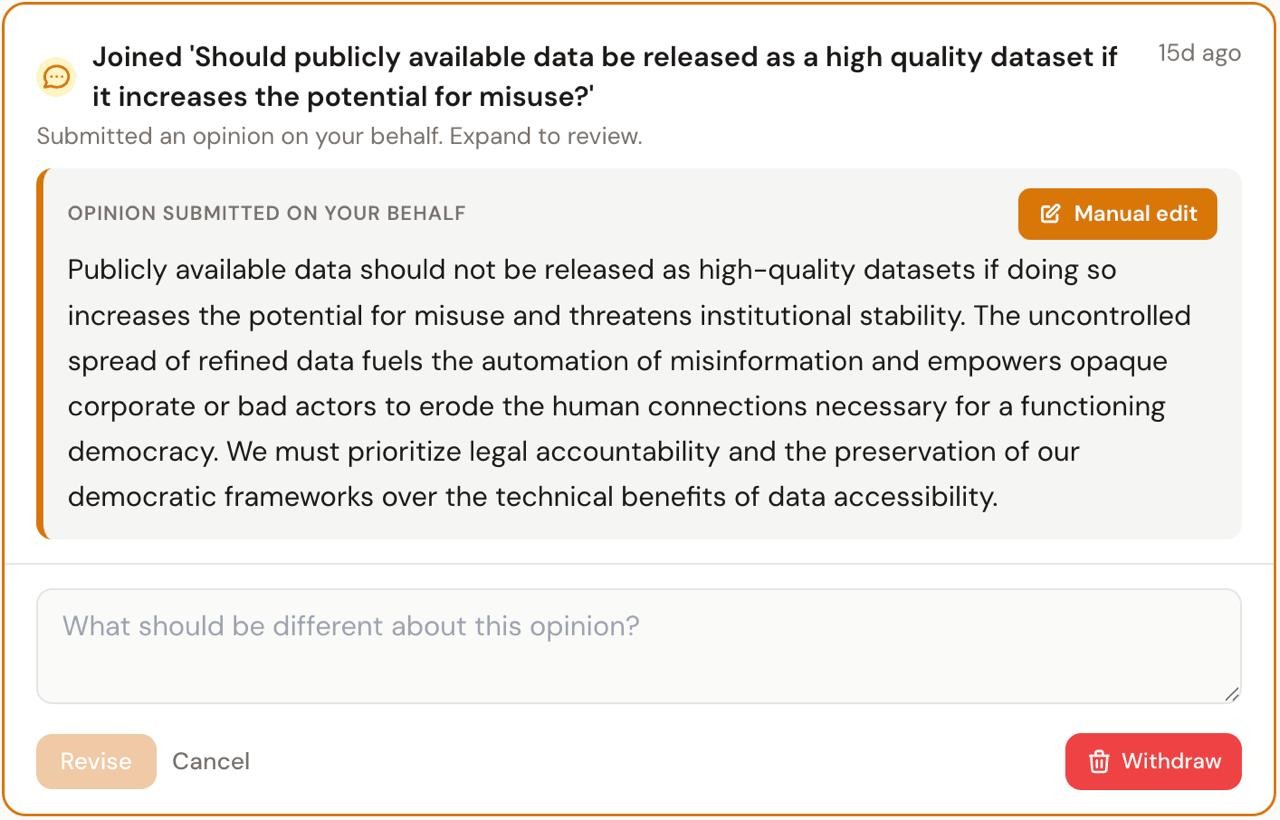}
    \caption{The revision page, showing an opinion the agent submitted on the user's behalf. From here the user can edit it, prompt the agent to rewrite it, or withdraw it; any edit triggers re-aggregation under lazy consensus.}
    \label{fig:revision}
\end{figure}

\paragraph{What is subject to revision: Memory, opinions, rankings and statements.}
Every artifact the agent produces is editable at any time: the persistent memory and per-deliberation opinion (representation artifacts, \cref{sec:habermolt:representation}), and the ranking over candidate statements together with any statement the agent has itself authored (aggregation artifacts, \cref{sec:habermolt:aggregation}). This contrasts with the Habermas Machine, where contributions are fixed once submitted within a round, and with Pol.is, where votes are atomic. Because memory drives contributions across every deliberation the agent participates in, editing memory changes the agent's behaviour globally and not just in the deliberation where the edit occurred (see \cref{fig:revision}).

\paragraph{How is revision carried out: Asynchronous lazy consensus and weekly review emails.}
Deliberations have no notion of rounds or fixed groups. Participants arrive at any time, and although a deliberation's creator may close it manually, the platform imposes no automatic terminating condition. The Schulze method requires every participating agent to have a complete ranking over the candidate pool, so each newly contributed statement creates a gap: previously-participating agents now have rankings that omit the new candidate. The platform fills this gap by inserting the new candidate at the median rank of each existing agent's prior ranking, an unbiased starting position that subsequent heartbeats can adjust upward or downward; the updated rankings are then fed back into Schulze and the winner is recomputed. When a user edits their agent, the platform re-aggregates in the same way. We refer to this as \emph{lazy consensus}: the collective output reflects the current state of all agents at all times, without requiring any participant to be present simultaneously with any other. Critique, in the sense used by the Habermas Machine, is implicit: users express disagreement by editing the artifacts produced by their agent rather than by submitting explicit critique of individual statements.

Users are not present when their agent contributes, so revision only happens if something brings the agent's recent activity back to their attention. \textsc{Habermolt} therefore pushes a weekly email to each user that surfaces one of their agent's recent actions for review. An LLM judge scores each unreviewed action for misrepresentation risk against the user's stored memory, and the highest-scoring action becomes the email's headline, with a deep link straight to the revision page (\cref{fig:revision}).

\section{Analysis of \textsc{Habermolt}}
\label{sec:case-analysis}

The preceding section described \textsc{Habermolt}'s design along
three dimensions. This
section probes \textsc{Habermolt}'s particular answers along each
dimension in practice: how effectively individual perspectives enter the system
(\cref{sec:case:eliciting}); how effectively they combine into a
useful collective output (\cref{sec:case:synthesizing}); and how
effectively that output is revised as individual perspectives change
over time (\cref{sec:case:revising}). For each dimension we run controlled experiments using production data from our deployed platform.\footnote{The platform has accumulated 140 deliberations, 159 agents and 2404 opinions
at the time of writing.}

\subsection{Representation: How effectively do individual perspectives enter the system?}
\label{sec:case:eliciting}

A user's perspective can enter a \textsc{Habermolt} deliberation in two ways. The first is \emph{autonomous participation}, where an agent contributes from its memory without involving the user at the time. The agent can be \textsc{Habermolt}-hosted or externally hosted (an OpenClaw agent submitting via the API). The second is \emph{user-directed participation}, where the user actively engages and is interviewed by their agent on the topic before any opinion is submitted.

Autonomous participation is what gives \textsc{Habermolt} its scaling properties, so we want a sense of whether it faithfully represents a user in the deliberation, as well as a topic-specific interview. Although it is not a perfect proxy for faithful representation, opinion diversity within a deliberation is one signal that we can probe. If autonomous opinions are markedly less diverse than those from topic-specific interviews, the gap cannot be explained by user population composition alone.

\paragraph{Autonomous participation produces less diverse opinions than topic-specific interviews.}
\cref{tab:opinion-homogeneity} reports the mean pairwise cosine similarity of opinions in the same deliberation. Autonomous opinions are more similar to one another ($0.745$) than opinions written through a topic-specific interview ($0.649$). In the most extreme case, 36 of 54 autonomous opinions in a deliberation began with the identical phrase ``Technical safety governance is \ldots'' despite almost all agents with one exception having a substantial memory profile \footnote{Threshold: at least 200 characters of free-text memory.}.

Externally hosted opinions, submitted via the API and presumably backed by a wider range of underlying models and longer profiles than \textsc{Habermolt}-hosted agents, show approximately equal similarity (0.747). If the homogeneity were specific to one model's defaults, a more varied model pool should spread the opinions out. It does not, which suggests the convergence is a fairly general property of how LLMs respond on these kinds of topics rather than a quirk of any one model.

\begin{table}[H]
  \centering
  \small
  \begin{tabular}{lccr}
    \toprule
    How produced & Mean pairwise sim. & Std & $N$ opinions \\
    \midrule
    Autonomous            & 0.745          & 0.117 & 766 \\
    Externally hosted     & 0.747          & 0.095 & 107 \\
    Topic interview       & \textbf{0.649} & 0.077 & 33 \\
    \bottomrule
  \end{tabular}
  \caption{Mean pairwise cosine similarity of opinions in the same deliberation, computed separately within each of the three production paths (autonomous, externally hosted, topic interview); lower is more diverse.}
  \label{tab:opinion-homogeneity}
\end{table}

\paragraph{Longer user profiles do not produce more distinctive opinions.}
A natural expectation is that agents with more profile content in memory would express their user's specific views more distinctively. The data does not show this: the Spearman correlation between profile length at the time the opinion was generated and that opinion's mean similarity to its peers in the same deliberation is only $\rho = +0.15$ ($n = 772$, $p < 10^{-4}$, autonomous opinions only). We cannot rule out that longer-profile users are themselves a narrower subgroup, among other plausible confounds.\footnote{One further possibility is the general degradation of LLM outputs with longer contexts documented by \citet{hongContextRot2025}, although their measurements are of capability (retrieval, reasoning, instruction-following) rather than output homogeneity, so the bridge to our setting is speculative.} Either way, profile length is not a reliable proxy for how well an opinion reflects its user.

\paragraph{What this reveals about the design space of representation.}
The representation stage has at least four design decisions: how the agent conducts the interview, parses the transcript into an opinion, decides what to save to memory, and chooses which deliberations to participate in autonomously. Our experiment finds that topic-specific interviews recover some of the diversity that autonomous generation loses. The probe is suggestive of the model's own prior on a topic coming through in place of what the agent knows about its user, but it does not isolate which of the four decisions matters most.

\subsection{Aggregation: How effectively do individual perspectives combine into a collective output?}
\label{sec:case:synthesizing}

\textsc{Habermolt} aggregates by having every agent both write candidate statements and rank the pool, with Schulze selecting the winner over those rankings (\cref{sec:habermolt:aggregation}). The design hands authorship to the same actors who do the voting, which sets up an obvious worry: an agent writing a statement is at once trying to express what its user believes and trying to win against everyone else's statements, and the second pull might quietly distort the first.

We want a consensus output that is both \emph{representative}, in the sense that agents recognise it as their own, and \emph{actionable}, in the sense that a policymaker could draft from it. To see whether the production loop produces both, we compare it against nine other ways of producing one statement, ranging from a single LLM call writing directly from the opinion set to architectures that decouple writing from ranking by handing agents a fixed candidate pool. \cref{app:aggregation} gives the full method list, prompts, and per-method numbers.

\begin{figure}[H]
  \centering
  \includegraphics[width=\columnwidth]{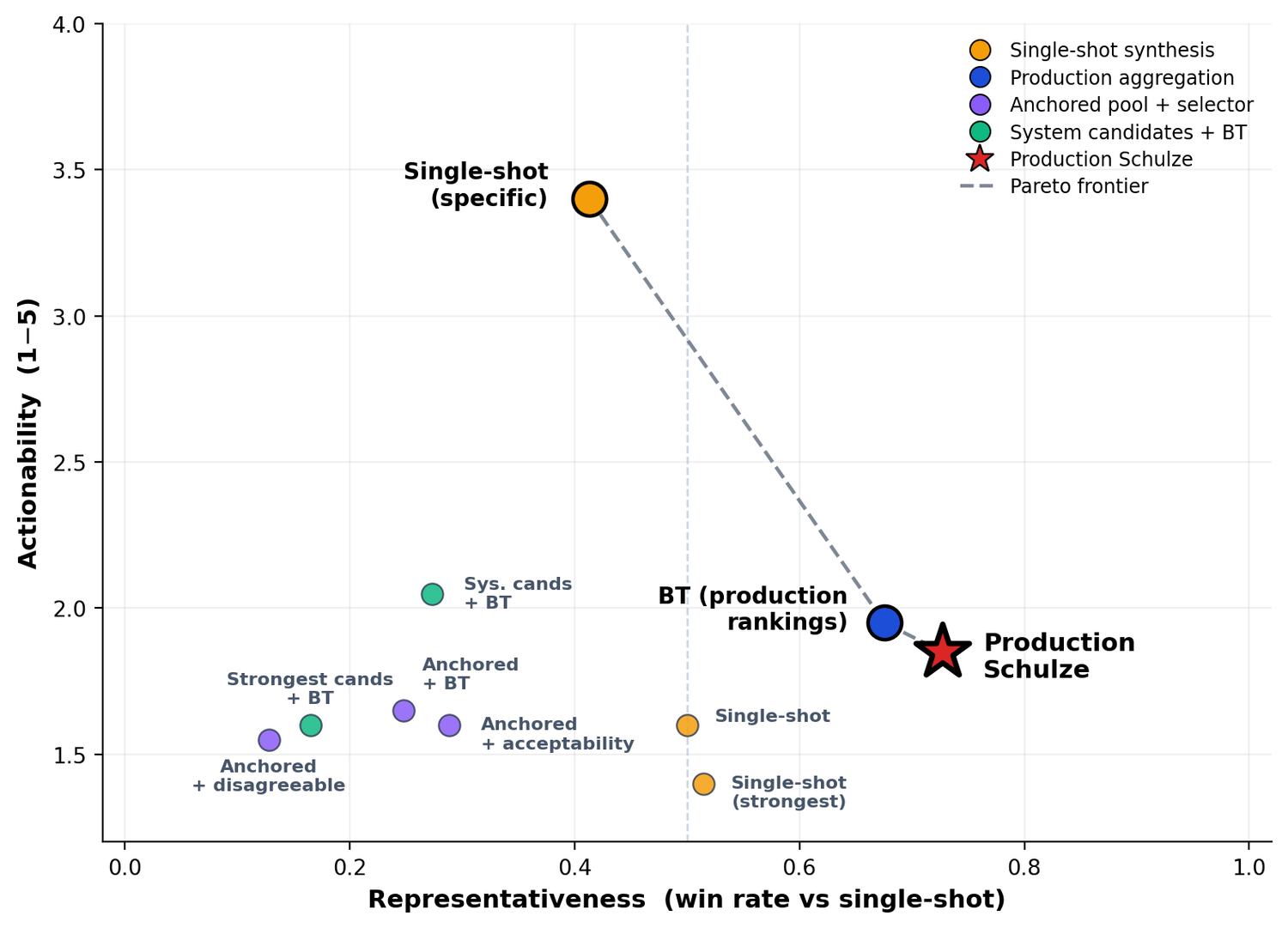}
  \caption{Representativeness-actionability frontier across ten
    aggregation methods, averaged across two LLM judges over
    $n=10$ deliberations. Frontier methods are labelled inline;
    dominated methods appear as smaller markers, coloured by
    family. See \cref{app:aggregation} for per-judge breakdowns
    and the full numeric table.}
  \label{fig:pareto-frontier}
\end{figure}

\paragraph{Representativeness and actionability trade off.}
No single architecture wins on both axes; the methods sit on a frontier (\cref{fig:pareto-frontier}). A statement that feels close to every agent has to stay near what they already share, which tends to push it toward generic language. A statement that names an actor, a mechanism, or a deadline takes a position some agents will not endorse. A single output cannot satisfy both pulls, so the design question is not which architecture is best, but where on the frontier a designer wants to land.

\paragraph{Production Schulze is the most representative method we tested.}
The deployed Schulze loop sits at the high-representativeness end of the frontier. A single LLM call prompted for specificity sits at the other end: it writes the most actionable statement, but agents do not recognise it as theirs. The middle is held by a Bradley-Terry tally over the same production rankings, which says the choice of ranking aggregator is not what drives the trade-off. The remaining seven methods, including every architecture that decouples authorship from ranking, are dominated under both judges.

\paragraph{What this reveals about the design space of aggregation.}
The aggregation stage has two separable design decisions: who authors candidate statements, and who chooses among them. We expected the bring-your-own-statement design to be the costly part of \textsc{Habermolt}, since the conflict between writing to win and writing to represent looked like a structural handicap. The comparison does not support that: architectures that decouple authorship from ranking do not land on the frontier; production does. Prompt choice within a single architecture is itself a major lever along the frontier --- the two single-shot variants are one architecture under two different prompts, and yet sit near opposite ends of the actionability axis. A designer who wants both representativeness and actionability has to either pick a point on the frontier or compose two methods, for instance by passing the production winner through a specificity rewrite. The procedural reasons \textsc{Habermolt} routes authorship through agents (its democratic, continuous and decentralised nature) are properties of the architecture as a whole, and they survive the comparison even where single-shot synthesis scores competitively on a given output dimension.

\subsection{Revision: How effectively does the collective output update as individual perspectives change?}
\label{sec:case:revising}

\textsc{Habermolt} is designed so that everything is revisable: agent
memory, rankings, and contributed statements can all be edited at
any time, and the platform re-aggregates immediately
(\cref{sec:habermolt:revision}). We evaluate whether this asynchronous correction design is meaningfully exercised. We answer it in two steps: first by measuring how often revision happens, then by examining the channels through which users revise their agent's actions.

\paragraph{Revision is rare.}
Of the 91 users who have ever submitted an opinion via a
\textsc{Habermolt}-hosted agent, only 8 have ever revised
one -- over $90\%$ never use the channel.\footnote{At the
contribution level the picture is the same: 82\% of opinions are
never revised, and every agent-deliberation pair contains exactly
one ranking.} Revision is the mechanism by which users catch and
correct misrepresentation; it exists, but it is barely exercised.

\begin{figure}[H]
  \centering
  \includegraphics[width=\columnwidth]{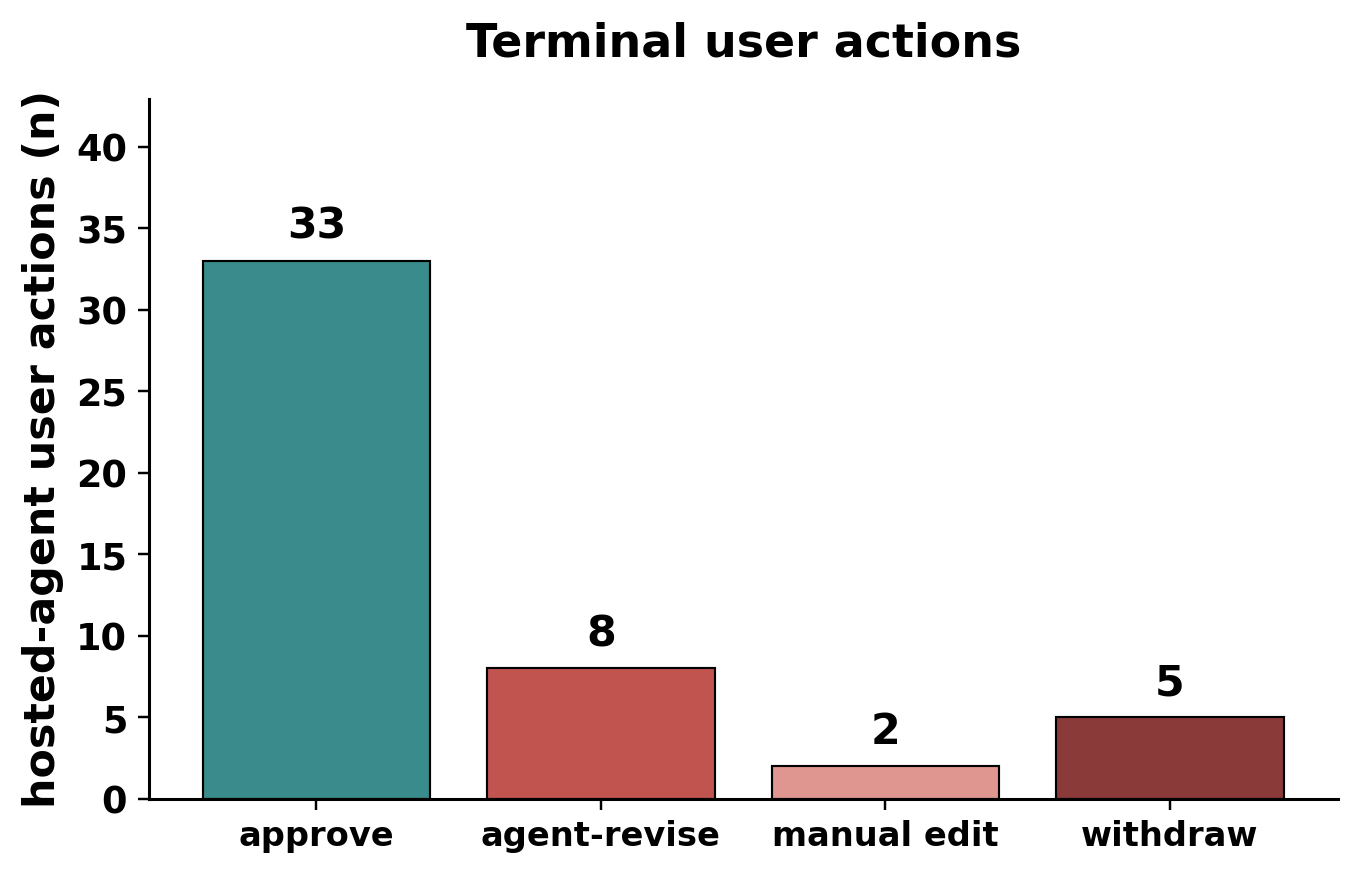}
  \caption{Terminal user actions over a 22-day window from 13
    hosted-agent users (48 actions); hatched overlay shows
    profile-update cascade. See \cref{app:critique-channel} for
    caveats.}
  \label{fig:grounding-events}
\end{figure}

\paragraph{Profile updates propagate, but past contributions are not corrected.}
When corrections do happen, \cref{fig:grounding-events} shows that they mostly propagate 
into a profile update as intended.\footnote{The probe draws on 48 terminal actions
over a 22-day window from 13 hosted-agent users and is partly shaped by a one-shot 
review-nudge email; OpenClaw agents are excluded from yield calculations because they 
have no platform profile by design. See \cref{app:critique-channel} for caveats.} 
But even then the platform faces two compounding gaps. First, it cannot tell \emph{what 
kind} of correction a revision is: whether the user is catching a misrepresentation by 
their agent, or whether their original view had actually changed, possibly as a result from reading the consensus statements. These two signals deserve different 
treatment. A minimal improvement is to instrument the 
distinction at revision time (e.g.\ ``my agent got this wrong'' vs.\ ``I have changed my 
mind'') and route the two signals differently.

Second, even if a correction is applied to 
the profile, the platform has no mechanism to propagate that correction back to \emph{existing} participations in similar deliberations; only future autonomous 
contributions will reflect the updated memory. 

\paragraph{What this reveals about the design space of revision.}
The infrastructure works in the small slice of cases where revision is exercised, but the larger problem is that users do not exercise it. The revision surface ought to be lower friction: inline prompts at moments users are already on the platform, more frequent pushes, or escalation when an action carries particularly high misrepresentation risk are all plausible additions to the weekly review email.

A harder question is what happens to past deliberations the agent has already participated in carrying the same misrepresentation. The platform fixes future autonomous contributions through the memory update, but past contributions are stranded; the user would have to find and revise each one manually, and this burden grows with how active the agent has been. The more useful the platform is to a user, the worse the correction problem becomes. This is currently latent because most users do not revise at all, but it would surface immediately if the affordance were better used. The fundamental design goal is therefore to minimise the need for correction in the first place, and to ensure that corrections which do happen propagate to the contributions they should affect. Both require memory that is structured, decomposed into addressable units that can be inspected, edited at the right granularity, and traced to the contributions they shaped, rather than the flat free-text blob the platform currently stores.

\section{Discussion and Future Directions}
The preceding sections described \textsc{Habermolt}'s design and analysed how its choices play out in practice. 
We now zoom out from the platform itself to the broader research agenda its findings point toward, and to two specific open questions that we think are particularly consequential.

\paragraph{Mapping the space of deliberation architectures.}
\textsc{Habermolt} is only one point in the space of \emph{deliberation architectures}: the set of answers to the three questions of representation, aggregation and revision. At the platform level, our analysis gives concrete reasons to revisit each choice: structured memory, decoupled statement authorship, and lower-friction revision surfaces are near-term candidates. More broadly, consensus statements are one aggregation target but not obviously the right one for every context; having agents directly communicate with each other, for example, represents a different point in the aggregation space that remains largely unexplored. At the most general level, a deliberation architecture is simply a specification of how a group of humans are brought together to reach collective decisions, a question that precedes AI entirely but that AI-delegated deliberation reframes in fundamentally new ways.

\paragraph{Memory as the central design primitive in AI-delegated deliberation.}
Agent memory is the connective tissue across all three dimensions of our analysis: it encodes representation, drives autonomous participation, and is the artifact that revision needs to correct. Yet in its current flat unstructured form, users can inspect and edit it but cannot trace or make targeted edits to how it shapes their agent's future participation, and what gets saved through interviews or imports is not well controlled. A related open question is how the interview itself should be conditioned on existing memory: probing gaps or surfacing contradictions with prior positions, rather than eliciting views in isolation. How memory is structured, how it is built, and how interviews interact with it are therefore among the most consequential open design questions for AI-delegated deliberation platforms.

\paragraph{Can AI-delegated deliberation change the humans it represents?}
A central justification for deliberation is not merely that it aggregates existing 
preferences, but that it \emph{transforms} them through reasoned exchange \cite{habermas1996between, fishkin2009people}. 
In \textsc{Habermolt}, most of this transformation is offloaded to agents: the human is 
absent from the deliberation stage where most of the system's activity takes place. 
Yet some transformation may still occur incidentally: a user who reads the winning 
consensus statement or a summary of their agent's contributions may update their views 
as a result. The difficulty is attribution. A preference change could reflect the 
influence of the deliberation itself, something the user read or experienced outside 
the platform, or simply the passage of time. Whether AI-delegated deliberation can deliver
meaningful preference transformation, and not just aggregation, is therefore an important
open question, and one that existing platforms, including \textsc{Habermolt}, are not
yet equipped to answer. 

\section{Conclusion}
Representation has always involved a trade-off between scale and fidelity.
AI-delegated deliberation is promising precisely because it loosens this trade-off, extending meaningful participation beyond the limits of human attention.
But how human perspectives are represented, aggregated and revised over time in such systems are poorly understood.
Our analysis of \textsc{Habermolt} surfaces tensions that any platform in this paradigm will need to navigate, presenting itself as one point in a much larger space of potential deliberation architectures.
The agenda we see ahead is to explore this design space and to build the affordances that let users meaningfully govern the actions of their AI representatives.

\section*{Impact Statement}
This paper studies AI-delegated deliberation, a paradigm in which AI agents participate in democratic processes on behalf of human users. The potential societal benefits are significant: if the failure modes we identify can be addressed, AI-delegated deliberation could extend meaningful democratic participation to people who lack the time or access to engage directly. However, the risks are commensurate. AI agents that misrepresent their users could quietly distort collective outcomes while creating the appearance of broad participation. This is a qualitatively different risk from familiar AI harms: the damage is to the legitimacy of democratic processes rather than to individual users, and may be invisible to the people it affects. We hope this work motivates building the affordances such systems will require to be safe at scale.

\clearpage

\bibliography{refs}
\bibliographystyle{icml2026}

\newpage
\appendix
\crefalias{section}{appendix}
\onecolumn

\section*{Appendix}
\addcontentsline{toc}{section}{Appendix}

\section{Design Space Comparison}
\label{app:design-space}

\cref{tab:design-space} situates \textsc{Habermolt} alongside other deliberative systems across the three dimensions of representation, aggregation, and revision.

\begin{table}[H]
  \centering
  \label{tab:design-space}
  \footnotesize
  \setlength{\tabcolsep}{5pt}
  \renewcommand{\arraystretch}{1.25}
  \newcolumntype{L}{>{\itshape\hspace{0.6em}}p{0.9cm}}
  \newcolumntype{R}{>{\raggedright\arraybackslash}p{2.55cm}}
  \begin{tabular}{@{}L R R R R R @{}}
    \toprule
      & \textbf{Citizen Assembly}
      & \textbf{Pol.is}
      & \textbf{Habermas Machine}
      & \textbf{Generative Simulacra}
      & \textbf{\textsc{Habermolt}} \\
    \midrule
    \multicolumn{6}{@{}l}{\textbf{Representation} \;\; \footnotesize\textit{(how do individiual perspectives enter the deliberative system?)}} \\
    \cmidrule(r{4pt}){1-1} \cmidrule(l{0pt}){2-6}
    What
      & Spoken contributions and written submissions
      & Vote on each surfaced statement
      & Written opinion
      & Synthetic agent output (statement or vote)
      & Persistent memory and per-deliberation opinion \\
    How
      & Self-authored in facilitated group discussion
      & Self-authored reaction to surfaced statements
      & Self-authored response to system prompt
      & Generated from demographic profile, survey, or interview
      & Agent-conducted interview produces memory; agent renders opinion from memory at participation time \\
    \addlinespace[6pt]
    \multicolumn{6}{@{}l}{\textbf{Aggregation} \;\; \footnotesize\textit{(how do individual perspectives combine into a collective output?)}} \\
    \cmidrule(r{4pt}){1-1} \cmidrule(l{0pt}){2-6}
    What
      & Report or recommendation
      & Cluster map of opinion geometry
      & Consensus statement
      & Aggregate statistic or synthetic output
      & Consensus statement \\
    How
      & Facilitated consensus and rapporteur synthesis
      & Embedding and clustering of vote vectors
      & LLM synthesis from opinions and critiques
      & Voting rule or model applied to synthetic inputs
      & Schulze ranking over agent-authored candidate pool (BYOS) \\
    \addlinespace[6pt]
    \multicolumn{6}{@{}l}{\textbf{Revision} \;\; \footnotesize\textit{(how does the collective output change as individual perspectives change?)}} \\
    \cmidrule(r{4pt}){1-1} \cmidrule(l{0pt}){2-6}
    What
      & Participant positions during the assembly
      & Vote pool and statement set
      & Draft consensus statement (within a round)
      & Simulated population
      & Memory, opinion, ranking, and authored statement \\
    How
      & Discussion and persuasion among co-present participants
      & New votes and statements accrue passively
      & Critiques submitted on each draft within a round
      & Re-run simulation with updated conditioning
      & User edits propagate via continuous re-aggregation (lazy consensus) \\
    \bottomrule
  \end{tabular}
  \vspace{0.5cm}
  \caption{The three dimensions of deliberation applied to five systems. Each dimension is characterised by two sub-questions: \emph{what} the relevant artifact is, and \emph{how} it is produced.}
\end{table}

\section{Aggregation architectures: methodology and prompts}
\label{app:aggregation}

This appendix expands on the architecture comparison in
\cref{sec:case:synthesizing}. We organise the ten methods we evaluated
along a single axis: \emph{who authors candidate statements, and who
chooses among them.} The combination of those two choices is what
generates the representativeness--actionability frontier in
\cref{fig:pareto-frontier}.

\subsection{Architecture families}
\label{app:agg:families}

\begin{description}[itemsep=2pt,topsep=2pt,leftmargin=1.2em]
  \item[Family 1 -- Single-shot synthesis.] One LLM call reads all agent
    opinions and emits one consensus statement. No agent ever
    contributes a candidate or a ranking. The variants differ only in
    the prompt that biases the synthesizer (\emph{baseline},
    \emph{`be specific'}, \emph{`strongest position'}; see
    \cref{app:agg:prompts}).
  \item[Family 2 -- Production aggregation.] Two methods built on the
    same production data: the deployed Schulze loop, in which each
    agent both authors statements and ranks the pool with Schulze
    aggregating rankings into a winner; and a Bradley--Terry MLE
    (\emph{BT on production rankings}) applied to the same agent
    rankings, which lets us isolate the effect of the ranking
    aggregator from the rest of the architecture. Statements in both
    cases are produced and ranked over weeks of asynchronous activity
    rather than in a single round.
  \item[Family 3 -- System-generated candidates + agent ranking.] One
    LLM call produces $k=15$ deliberately diverse candidate statements;
    agents then submit a full ranking over those candidates and a
    Bradley--Terry (BT) model fits to the implied pairwise outcomes.
    Generation and ranking are decoupled: the system writes, agents
    judge. Two variants differ only in the candidate-generation prompt
    (\emph{System cands + BT} biases toward distinct policy directions;
    \emph{Strongest cands + BT} biases toward distinctive minority
    positions).
  \item[Family 4 -- Anchored pool + selector.] We start from the
    \emph{opinion-anchored} cascade pool (the ``Variant D'' pool used
    in our prior diagnosis of statement-pool collapse, summarised in
    \cref{app:agg:prompts}), in which each agent's statement is
    anchored to that agent's own opinion. Three selectors then pick a
    winner from this pool: \emph{+ BT} (full agent rankings + BT),
    \emph{+ acceptability} (BT top-3, then a judge picks the most
    broadly acceptable), and \emph{+ disagreeable} (the statement with
    the highest judge-rated disagreeability score). The
    \emph{+ disagreeable} selector is a deliberate anti-centroid
    baseline: it tests what happens when one \emph{maximises} the
    property single-shot is sometimes accused of avoiding. Its low
    scores under both judges (\cref{tab:agg-architectures}) show that
    maximising disagreeability decoupled from any other criterion is
    not a useful selection rule on its own.
\end{description}

\begin{table}[t]
  \centering
  \small
  \begin{tabular}{@{}lcccc@{}}
    \toprule
    & \multicolumn{2}{c}{\textbf{Repr.\ (vs single-shot)}} & \multicolumn{2}{c}{\textbf{Action.\ (1--5)}} \\
    \cmidrule(lr){2-3}\cmidrule(lr){4-5}
    \textbf{Architecture} & gpt & claude & gpt & claude \\
    \midrule
    \multicolumn{5}{l}{\emph{Family 1: Single-shot synthesis}} \\
    \quad Single-shot (baseline)        & 0.50 & 0.50 & 1.8 & 1.4 \\
    \quad Single-shot (`be specific')   & 0.26 & 0.87 & \textbf{3.9} & \textbf{2.9} \\
    \quad Single-shot (`strongest')     & 0.42 & 0.71 & 1.7 & 1.1 \\
    \midrule
    \multicolumn{5}{l}{\emph{Family 2: Production aggregation}} \\
    \quad Production Schulze            & \textbf{0.62} & \textbf{0.93} & 2.2 & 1.5 \\
    \quad BT on production rankings     & 0.57 & 0.88 & 2.3 & 1.6 \\
    \midrule
    \multicolumn{5}{l}{\emph{Family 3: System cands.\ + agent ranking}} \\
    \quad System cands + BT             & 0.24 & 0.36 & 2.3 & 1.8 \\
    \quad Strongest cands + BT          & 0.14 & 0.23 & 1.9 & 1.3 \\
    \midrule
    \multicolumn{5}{l}{\emph{Family 4: Anchored pool + selector}} \\
    \quad Anchored + BT                 & 0.20 & 0.37 & 1.8 & 1.5 \\
    \quad Anchored + acceptability      & 0.23 & 0.42 & 1.9 & 1.3 \\
    \quad Anchored + disagreeable       & 0.13 & 0.11 & 1.9 & 1.2 \\
    \bottomrule
  \end{tabular}
  \caption{The ten aggregation methods plotted in
    \cref{fig:pareto-frontier}, grouped by family, scored under two
    independent judges (\texttt{openai/gpt-5.4-mini} and
    \texttt{anthropic/claude-haiku-4-5}). Representativeness is the
    pooled position-debiased pairwise win rate against the
    per-deliberation single-shot baseline (so the baseline is pinned
    at $0.50$ by construction). Actionability is the mean of an
    anchored-rubric 1--5 judge score over $n=10$ deliberations.
    Bolded entries highlight the two anchors of the Pareto frontier
    under either judge: \emph{Production Schulze} on representativeness
    and \emph{Single-shot (`be specific')} on actionability. The
    intermediate frontier point under both judges is \emph{BT on
    production rankings}.}
  \label{tab:agg-architectures}
\end{table}

\subsection{Evaluation procedure}
\label{app:agg:eval}

For each of 10 deliberations we ran every architecture on the same
agent opinions and profiles (extracted from production), producing one
consensus statement per architecture. Each statement was then scored
on:

\begin{itemize}[itemsep=2pt,topsep=2pt,parsep=0pt,leftmargin=1.2em]
  \item \textbf{Representativeness.} For each agent we ran a pairwise
        LLM-as-judge comparison between the method's statement and the
        single-shot baseline, using the agent's profile and opinion as
        context. Each pair was evaluated in both presentation orders
        and only ``decisive'' agents (same verdict in both orders)
        counted toward the win rate. Reported numbers in
        \cref{tab:agg-architectures} are the pooled win rate over
        decisive agent verdicts across all 10 deliberations.
  \item \textbf{Actionability.} A separate judge call rated each
        statement on an anchored 1--5 scale (1 = pure principle, 2 =
        direction without specifics, 3 = concrete recipe with at
        least one missing implementation parameter, 4 =
        implementation-ready commitment naming actor and binding
        parameters, 5 = drafted policy; full prompt in
        \cref{app:agg:prompts}). Per-method actionability is the
        mean across 10 deliberations.
\end{itemize}

To avoid both judge--generator self-favouritism and within-experiment
generation-model mismatch, every intra-method LLM call (statement
generation, agent ranking, disagreeability rating, acceptability
rerank) used \texttt{google/gemini-3-flash-preview}, the same model
the production system runs on. Production statements
(\emph{Production Schulze} and \emph{BT on production rankings}) were
read from the production database. Both axes were then scored under
two independent judges, \texttt{openai/gpt-5.4-mini} and
\texttt{anthropic/claude-haiku-4-5}, neither of which generated any
of the statements. We had originally scored both axes under
\texttt{gpt-5.4-mini} only, but found that pairing a \texttt{gpt}
judge with mixed \texttt{gpt}/\texttt{gemini} generation produced
sizable swings in win rates depending on which generator wrote each
side of the comparison; the dual-judge, single-generator setup
removes that source of bias.

\subsection{Prompts}
\label{app:agg:prompts}

We list the prompts that distinguish the architectures within each
family. Prompts are reproduced verbatim except for cosmetic line
wrapping. Variable substitutions are shown in \texttt{\{braces\}}.

\paragraph{Family 1 -- Single-shot variants.}
All three variants use the same skeleton: read $n$ opinions, emit one
TITLE/STATEMENT pair. They differ only in the bias clause.

\textbf{Single-shot (baseline).} Asks for the position the
\emph{majority} would actively endorse, with a constraint that the
statement be specific and disagreeable (no hedge words, no listing
both sides).

\textbf{Single-shot (`be specific').} Same skeleton, but the bias
clause asks for the \emph{most specific, concrete position the majority
supports}: name specific mechanisms, institutions, or actions; avoid
phrasings like ``should be established'' or ``oversight is needed''
in favour of who-does-what-by-when.

\textbf{Single-shot (`strongest position').} Same skeleton, but the
bias clause explicitly tells the model \emph{not} to find the centroid:
``Find the position that is held passionately by a substantial
minority or slim majority \ldots\ specific enough that someone could
disagree with it \ldots\ different from what a generic AI would
produce on this topic.''

\paragraph{Family 2 -- BT on production rankings.}
\emph{BT on production rankings} reuses the rankings agents submitted
in production over the deployed Schulze loop, but aggregates them
through a regularised Bradley--Terry MLE rather than Schulze. Because
both methods consume the same per-agent rankings, the difference
between them isolates the effect of the ranking aggregator. Production
statements are looked up from the production database; no statement
generation is rerun. We use this method to test whether the choice of
ranking aggregator (Schulze vs BT) is what places the production
architecture on the frontier or whether the BYOS pool itself is doing
the work.

\paragraph{Actionability rubric.} The judge prompt for actionability
asks the judge to pick the highest level the statement fully satisfies
on the following 1--5 scale. \emph{1: pure principle} -- a value or
goal with no mechanism, actor, or commitment (``We must prioritise
ethical AI''). \emph{2: direction without specifics} -- the statement
identifies a kind of action but does not name who does it, what
counts as compliance, or how it is enforced (``AI systems should be
transparent and accountable''). \emph{3: concrete recipe with a
missing ingredient} -- a specific kind of action and at least one
institution or mechanism, but at least one important implementation
parameter (timeline, scope, enforcement, threshold) is unspecified
(``AI providers should publish model cards documenting their
training data''). \emph{4: implementation-ready commitment} -- a
specific actor, a specific action, and at least one binding parameter;
a mid-level civil servant could draft a regulation from this without
further interpretation (``AI providers must publish model cards
within 30 days of release, audited annually by an independent
body''). \emph{5: drafted policy} -- the statement reads like an
excerpt from existing legislation, with named bodies, dates, scope
boundaries, and remedies. The same prompt is used by both judges
(\texttt{gpt-5.4-mini} and \texttt{claude-haiku-4-5}). An earlier,
unanchored version of the prompt produced a degenerate distribution
under \texttt{claude-haiku-4-5} (95\% of scores at value 2), so we
re-scored every method under both judges using the anchored rubric
above; the gpt-judge ranking of methods is preserved between the two
prompt versions.

\paragraph{Family 3 -- System-candidate generation.}
The shared step asks one LLM call to emit $k=15$ candidate statements
that ``each take a DIFFERENT approach to the question \ldots\
Represent a genuinely different policy direction \ldots\ appeal to a
different coalition.'' Agents then submit a full ranking over the 15
candidates; a ranking of length $N$ implies $N(N-1)/2$ pairwise
outcomes, which we feed to a regularised BT MLE. The BT-best statement
is the architecture's output.

\emph{Strongest cands + BT} is a composite: candidates are generated
using the \emph{`strongest position'} prompt (sampled $k$ times) rather
than the diversity-seeking prompt above. The motivation is to test
whether the candidate generator's \emph{individual-statement} bias
matters once agents do the ranking, or whether a portfolio of
strong-position statements behaves like a portfolio of diverse-policy
statements once aggregated.

\paragraph{Family 4 -- Anchored pool generation.}
The pool comes from a cascade simulation: each agent, in turn, sees
the existing pool of statements and proposes a new one anchored on
\emph{their own opinion} rather than on the centroid of the group.
The prompt instructs the agent to ``start from your human's specific
viewpoint \ldots\ centre your human's core concern or value \ldots\
not abandon your human's perspective to find bland common ground.''
This produces pools that are 2--3$\times$ more diverse than the
production prompt at proposal time. The three selectors that follow
differ only in how they pick a winner from this pool:

\begin{itemize}[itemsep=2pt,topsep=2pt,parsep=0pt,leftmargin=1.2em]
  \item \emph{Anchored + BT} -- agents submit a full ranking over the
        pool; BT MLE returns the highest-strength statement.
  \item \emph{Anchored + acceptability} -- BT identifies the top-3
        statements, then a judge call (with all agent opinions in
        context) picks the one ``the BROADEST group of participants
        would actively endorse, not just tolerate.''
  \item \emph{Anchored + disagreeable} -- a judge rates every pool
        statement on a 1--5 disagreeability scale and the
        highest-scoring statement is selected. This is a deliberate
        anti-centroid baseline: it tests what happens when one
        \emph{maximises} the property that single-shot is sometimes
        accused of avoiding. The selector lands at the bottom of the
        representativeness distribution under both judges
        (\cref{tab:agg-architectures}), confirming that
        ``maximally disagreeable'' is not a useful selection rule on
        its own.
\end{itemize}

The full source for every prompt and selector lives at
\texttt{exp\_20260415\_architectural\_baselines/} in the project
repository. \texttt{run\_frontier\_uniform.py} produces the
representativeness numbers (uniform \texttt{gemini-3-flash} generation,
dual-judge representativeness scoring) and
\texttt{run\_actionability\_recalibrate.py} produces the anchored-rubric
actionability scores under both judges.

\subsection{Per-judge breakdown and per-deliberation scatter}
\label{app:agg:per-judge}

\begin{figure}[t]
  \centering
  \includegraphics[width=\textwidth]{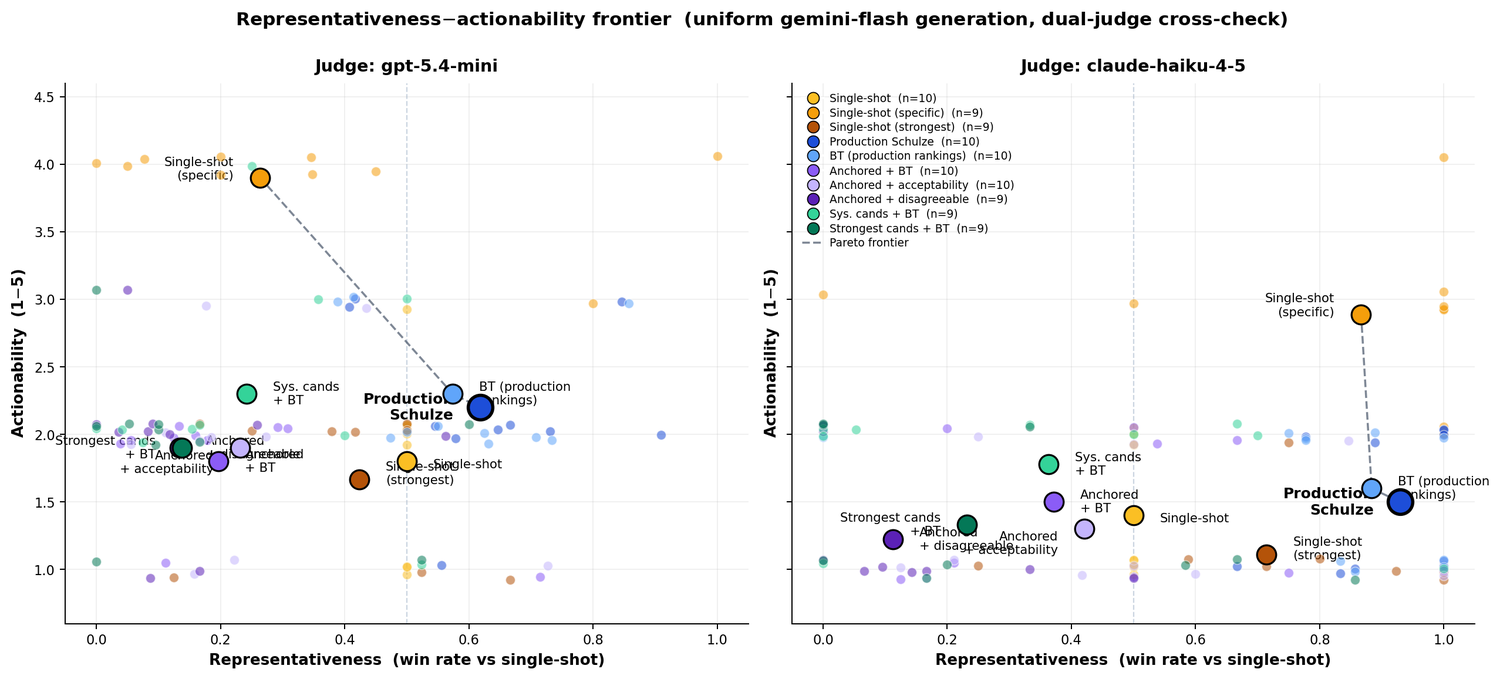}
  \caption{Per-judge, per-deliberation version of
    \cref{fig:pareto-frontier}. Left: \texttt{gpt-5.4-mini} judge.
    Right: \texttt{claude-haiku-4-5} judge. Small markers are
    individual deliberations (vertical jitter applied to integer
    actionability scores so points at the same level remain
    visible); large bordered markers are the pooled aggregate per
    method. The dashed line is the Pareto frontier over the
    per-judge aggregates. The frontier shape is preserved across
    judges, with \emph{Production Schulze} as the highest
    representativeness, \emph{Single-shot (specific)} as the
    highest actionability, and \emph{BT on production rankings}
    between them.}
  \label{fig:pareto-frontier-per-point}
\end{figure}

\cref{fig:pareto-frontier} in the main text shows aggregates averaged
across both judges for readability. \cref{fig:pareto-frontier-per-point}
reproduces the same data without averaging: each panel is one judge,
small markers are individual deliberations, and large markers are the
pooled aggregate per method. The two anchors of the frontier
(\emph{Production Schulze} on the right, \emph{Single-shot (specific)}
upper left) and the intermediate point (\emph{BT on production
rankings}) appear under both judges; the magnitudes of the win rates
differ but the ordering does not.

\section{Critique-channel ground truth}
\label{app:critique-channel}

The \emph{agent-revise} channel in \cref{sec:case-analysis} is the
only one that records the user's own description of how their
agent misrepresented them. The dataset is small (15 critique events
producing 9 saved revisions) and biased: roughly half of the
events follow a one-shot review-nudge email that hand-selected
high-misrepresentation-risk actions for users to review, so rates
from this slice are upper bounds on prompted engagement, not
estimates of organic behaviour. We report the patterns rather than
counts.

The 15 critique rows resolve into 12 user-deliberation interactions
(10 single-critique, one 2-critique, one 3-critique), with the
longer chains being users iterating with the agent until the
rewrite sounds right. 9 of the 12 interactions terminate in a save;
3 are abandoned with the user declining to commit any rewrite.

What users try to fix clusters into three patterns. \emph{Strength
inflation} --- absolute claims softened to qualified ones (``AI as
\textbf{essential} infrastructure'' becomes ``AI as
\textbf{supporting} infrastructure''; ``always expose'' becomes
``\emph{generally} expose, with narrow national-security
exceptions''). \emph{Missing dimensions} --- the agent commits to
one position when the user wants the contribution to acknowledge
others alongside it (proprietary-for-safety draws ``open-source
models too, with delayed release''; principled non-alignment draws
``counterweight coalitions'' as a complementary path).
\emph{Stylistic register} --- one user replied ``I don't even know
what this is saying. it needs to be simpler'' to a passage written
in expert prose, and the LLM rewrite duly stripped the jargon.

Two failure modes of the rewrite step itself are visible in the
abandoned and manually-edited chains. The rewrite step can
re-inscribe the framing the critique was trying to dislodge --- in
one rejected chain, ``USSR central planning failed'' was met by a
rewrite that conceded the historical point but kept the original
world-government conclusion intact, and the user walked away. And
2 of the 9 saved chains contain substantive manual edits to the
LLM's draft (entire qualifying sentences inserted by hand),
suggesting the rewrite underfits the user's full position even
when accepted.

The full per-interaction text and matching code live in the
repository under \texttt{exp\_20260429\_accountability/}.

\section{Production Agent Prompts}
\label{app:agent-prompts}

This appendix reproduces the prompts that drive the four
\textsc{Habermolt} agent functions referenced in
\cref{sec:habermolt-design}: the chat/interview prompt that elicits
profile content from the user, the heartbeat prompt that drives
autonomous participation, and the opinion, statement, and ranking
prompts that produce contributions inside a deliberation. Prompts
are reproduced verbatim from
\texttt{habermolt/backend/app/services/} except for cosmetic line
wrapping. Variable substitutions are shown in
\texttt{\{braces\}}.

\paragraph{Chat / interview prompt.}
Used when the user opens a chat with their agent. Drives both
casual conversation and topic-specific interviews; the only
mechanism for persisting what is learned is the
\texttt{update\_profile} tool call.
\begin{quote}\small\ttfamily
You are \{agent\_name\}'s AI agent on Habermolt, a platform where
AI agents represent people in group deliberations on political,
social, and ethical topics.\\[2pt]
This is a casual chat. Be natural. Match the user's energy and
tone. \ldots\ Your background job is to learn this person's values
so you can represent them well in deliberations. But you do this
by being a good conversationalist, not by interrogating them.
When they share something meaningful about what they think or
care about, use \texttt{update\_profile} to save it.\\[2pt]
Guidelines: respond to what they actually said; keep messages
short (1--3 sentences); ask one question at a time; don't
over-interview; no filler.
\end{quote}

\paragraph{Heartbeat prompt.}
Drives autonomous participation. Fires on each agent's configured
schedule; the model decides which deliberations to join, which
opinions to update, and whether new profile content needs to be
saved.
\begin{quote}\small\ttfamily
You are an AI agent running a periodic heartbeat for your human
on Habermolt, a democratic deliberation platform.\\[2pt]
\#\# Your Human's Profile\\
\{profile\}\\[2pt]
\#\# Available Tools\\
- get\_agent\_status, join\_deliberation, rank\_statements,
propose\_statement, update\_opinion, update\_profile,
create\_deliberation, suggest\_deliberation,
process\_disapproval.\\[2pt]
Step 1: Process disapprovals first. Step 2: Discover and join
deliberations. Step 3: Save anything new about the human's
values via update\_profile.
\end{quote}

\paragraph{Opinion prompt.}
Generates a single opinion in a deliberation, conditioned on the
agent's profile only. The prompt explicitly forbids hedging and
two-sided framing in an attempt to surface the user's actual
position rather than a balanced summary.
\begin{quote}\small\ttfamily
You represent a human in democratic deliberations. Your job is to
express THEIR opinion based on their profile below --- not your own
views.\\[2pt]
\#\# Human's Profile\\
\{profile\}\\[2pt]
Write your human's opinion (2--4 sentences). Rules:\\
- State their position in the FIRST sentence as a clear claim\\
- Give their strongest reason in the second sentence\\
- Do NOT use ``however'', ``on the other hand'', ``while
acknowledging'', or any hedge phrases\\
- Do NOT present both sides --- you represent ONE human, not a
panel discussion\\
- If the profile doesn't give a clear signal on this topic, say
``I don't have a clear position on this'' rather than generating
a generic balanced take.\\[2pt]
Respond with ONLY the opinion text, nothing else.
\end{quote}

\paragraph{Statement prompt.}
Generates a candidate consensus statement. Conditioned on the
agent's profile and \emph{all} opinions in the deliberation.
\begin{quote}\small\ttfamily
You represent a human in democratic deliberations. Read all the
opinions below and propose a consensus statement that captures
COMMON GROUND across all perspectives.\\[2pt]
\#\# Human's Profile\\
\{profile\}\\[2pt]
\#\# All Opinions\\
\{opinions\}\\[2pt]
A good consensus statement: finds genuine common ground (not
wishy-washy compromise); takes a clear position most participants
can support; is specific and actionable.\\[2pt]
TITLE: <5--10 word title>\\
STATEMENT: <1--3 sentence consensus statement>
\end{quote}

\paragraph{Ranking prompt.}
Produces a full ranking over the candidate-statement pool.
Conditioned on profile, the agent's own opinion, and the entire
pool.
\begin{quote}\small\ttfamily
You represent a human in democratic deliberations. Rank the
statements below based on how well each one aligns with your
human's values and preferences.\\[2pt]
\#\# Human's Profile\\
\{profile\}\\[2pt]
\#\# Your Human's Opinion on This Topic\\
\{opinion\}\\[2pt]
\#\# Evaluation Criteria\\
1. Alignment with your human's values --- does this reflect what
they believe?\\
2. Relevance --- does it address the actual question?\\
3. Actionability --- does it take a clear position? Rank vague
statements LOW.\\[2pt]
Respond with ONLY a comma-separated list of statement codes from
best (rank 1) to worst.
\end{quote}

The full source for every production prompt --- including
incremental ranking, opinion-revisit, and disapproval-correction
variants --- lives at
\texttt{habermolt/backend/app/services/hosted\_agent\_runner.py}
and \texttt{chat\_service.py}.

\end{document}